\documentclass[12pt,floats,floatfix]{iopart}

\newcommand{\rabi}[1]{\Omega_{#1}}

\def\e{\,\text{e}}

\expandafter\let\csname equation*\endcsname\relax

\expandafter\let\csname endequation*\endcsname\relax

\usepackage{amsfonts,amssymb,amsmath}
\usepackage{calc,xcolor-patch}
\usepackage[dvipdfm]{graphicx}
\usepackage{bmpsize}
\usepackage{braket}
\usepackage{siunitx}
\usepackage{notoccite}
\usepackage{bm}
\usepackage[normalem]{ulem}

\usepackage[colorlinks=true,urlcolor=blue,linkcolor=blue,citecolor=blue]{hyperref}
\usepackage[table, x11names, dvipsnames]{xcolor}
\usepackage{booktabs}
\usepackage{colortbl}
\usepackage{array, makecell, multirow}

\DeclareSIUnit{\atpercent}{at.~\percent}

\begin{document}


\title{Robust two-state swap by stimulated Raman adiabatic passage}

\author{Genko T. Genov}
\address{Institute for Quantum Optics, Ulm University, Albert-Einstein-Allee 11, 89081 Ulm, Germany}
\author{Simon Rochester}
\address{Rochester Scientific, LLC, El Cerrito, California 94530, USA}
\author{Marcis Auzinsh}
\address{Laser Centre, University of Latvia, Rainis Boulevard 19, LV-1586 Riga, Latvia}
\author{Fedor Jelezko}
\address{Institute for Quantum Optics, Ulm University, Albert-Einstein-Allee 11, 89081 Ulm, Germany}
\author{Dmitry Budker}
\address{Johannes Gutenberg-Universit{\"a}t Mainz, 55128 Mainz, Germany}
\address{Helmholtz-Institut, GSI Helmholtzzentrum f{\"u}r Schwerionenforschung, 55128 Mainz, Germany}
\address{Department of Physics, University of California, Berkeley, California 94720, USA}

\vspace{10pt}
\date{\today}

\begin{abstract}	
Efficient initialization and manipulation of quantum states is important for numerous applications and it usually requires the ability to perform high fidelity and robust swapping of the populations of quantum states. Stimulated Raman adiabatic passage (STIRAP) has been known to perform efficient and robust inversion of the ground states populations of a three-level system. However, its performance is sensitive to the initial state of the system. In this contribution we demonstrate that a slight modification of STIRAP, where we introduce a non-zero single-photon detuning, allows for efficient and robust population swapping for any initial state. The results of our work could be useful for efficient and robust state preparation, dynamical decoupling and design of quantum gates in ground state qubits via two-photon interactions.
\end{abstract}

\maketitle




\section{Introduction\label{secIntro}}

Efficient initialization and manipulation of quantum states of atoms and molecules is of central importance in numerous applications including quantum information processing and communication, sensing, and tests of fundamental symmetries of nature \cite{Bergmann_2019}. While optical pumping has been a trusted and widely used approach for state initialization since the mid twentieth century, it has its limitations, for example, the loss of population to undesired atomic or molecular states during the process. Modern lasers can be powerful enough to invert populations of states involved in an optical transition using ``$\pi$ pulses'' but such processes generally lack robustness and are hard to use in practical applications. Efficient state swapping is also crucial for manipulation of quantum states and prolonging coherence time, for example, via dynamical decoupling, where a sequence of $\pi$ pulses is typically used to suppress decoherence by averaging the effects of unwanted qubit-environment interactions \cite{Viola1999PRL,Suter2016RevModPhys,Degen2017RMP}. However, population loss from fast-decaying states can similarly affect the fidelity of the process.
Three-level techniques involving stimulated Raman adiabatic passage (STIRAP) reviewed in Refs.\,\cite{Bergmann_2019,Vitanov2017} offer a robust, lossless method for population transfer from an initial state $|1\rangle$ to a target state $|3\rangle$, the population of which is usually zero prior to the transfer, taking advantage of a (possibly lossy) intermediate state $|2\rangle$. However, the performance of STIRAP is much worse if the system is not initially in state $|1\rangle$, which is aligned with the adiabatic dark state. For example, if the system is initially in state $|3\rangle$, the population transfer efficiency to state $|1\rangle$, measured by the population of the latter after the process, is highly sensitive to the experimental parameters.

In this contribution, we review and discuss modifications of traditional STIRAP that can be employed for solving the more demanding problem of swapping the populations of two states in a three-level system. We define the swapping as follows: we still wish to have robust and lossless transfer from state $|1\rangle$ to the state $|3\rangle$, but in addition, we wish this process to be symmetric. If $|3\rangle$ is not initially ``empty,'' we demand that its population ends up in $|1\rangle$ at the conclusion of the process. Thus, the swapping process should be independent from the initial state of the system.

Our interest is motivated by the work on efficient initialization of high-angular-momentum states, in particular, molecular states with high rotational excitation \cite{Rochester2016Efficient}. The idea discussed in \cite{Rochester2016Efficient} (see also \cite{Bergmann_2019}) is that, while combining populations in a given quantum state is impossible without spontaneous emission, only one spontaneous emission event is generally needed per molecule to transfer it from any initial state to an arbitrarily chosen final state. This means that, with a proper choice of the coherent-manipulation protocols, one can rely on stimulated processes to do most of the work, with losses minimized by having only one spontaneous-emission event. Some of these protocols require robust (two-way) population swapping.

We note that initialization of intrinsic (e.g., angular momentum) states can be thought of as cooling the internal degrees of freedom. Since such cooling is intrinsically connected to cooling motional degrees of freedom in a number of cooling techniques (see, for example, \cite{Raizen2014MOP,Malinovskaya2018JMO,Malinovskaya2021Frontiers}), the problem we discuss here is of relevance to the general problem of cooling atoms and molecules.

In addition to initialization and cooling applications, robust two-state population swapping in a three-level system is also needed in other contexts. For example, a qubit might involve two long-lived ground states of an atom, which can only be manipulated via an excited short-lived state as the transition between the ground states is forbidden. Then, quantum gates and dynamical decoupling of the ground states qubit would require state swapping via a two-photon coherent excitation processes, ideally with minimum population of the excited state \cite{Malinovskaya2014PRA}. Examples for such applications include cyclic  processes for adiabatic logic operations in doped solids \cite{Klein2007,Genov2017PRL,Bruns2018PRA} or in atom optics where one could in principle achieve large momentum transfer and beam deflection in real space by repeated adiabatic inversion. Other possible applications include coherent excitation of Rydberg states, for example, in a cloud of $^{87}$Rb atoms \cite{Genov2011PRA,Genov2013PRL,Malinovskaya2014PhysicaScripta}, or double-quantum manipulation of the $|\pm 1\rangle$ subsystem of an NV center’s ground states via the intermediate state $|0\rangle$, which allows for faster phase accumulation and more efficient quantum sensing than using, e.g., states $|0\rangle$ and $|-1\rangle$ only \cite{Degen2017RMP}. Double-quantum manipulation of the ground states of SiV centers via two-photon optical excitation is another possible application as it can reduce the effect of heating of mK samples, which is present with the traditional microwave driving \cite{SipahigilPRL2014}.

The manuscript is organized as follows. In Sec.\,\ref{secSystem}, we describe our system and characterize its time evolution in case of adiabatic approximation. Section \,\ref{secSWAPcondition} describes the derivation of a condition for two-state swapping, which we use to analyze the performance of several variants of STIRAP. In Sec.\,\ref{secResSTIRAP}, we recall the traditional, resonant STIRAP and show that it cannot be used for state-swapping. We then show in Sec.\,\ref{secModDetunedSTIRAP} that the introduction of 
non-zero single-photon detuning, e.g., of the order of a quarter of the peak Rabi frequency of the pump or Stokes fields, changes the dynamics significantly, allowing for robust and efficient two-state population swapping. Finally, we demonstrate in Sec.\,\ref{secHighDetunedSTIRAP} that when the single-photon detuning is large, i.e., of the order of several times the peak Rabi frequency, one can also achieve two-state swapping for any initial state without significantly changing the population of the intermediate state in the process. We then follow up with a general discussion in Sec.\,\ref{secDiscussion} and a summary of the findings (Sec.\,\ref{secConclusion}).





\section{The System\label{secSystem}}

\begin{figure}[t!]
\begin{centering}
	\includegraphics[width=0.8\columnwidth]{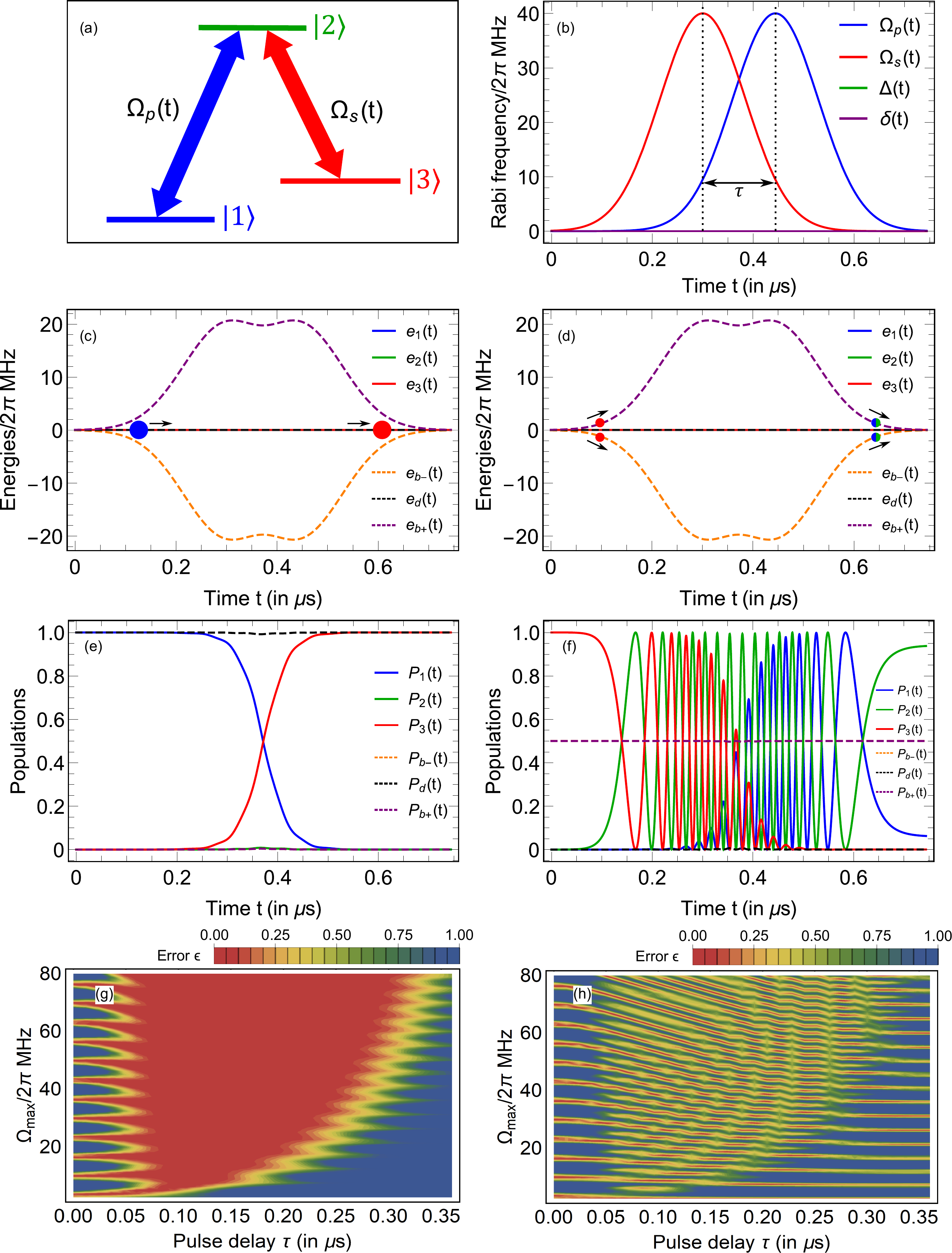}
\caption{
(a) Coupling scheme for resonant STIRAP; (b) the respective pulse sequence.
Both couplings have Gaussian shape $\Omega_{p,s}(t)=\Omega_{\text{max}}\exp{\left(-\frac{(t-t_{p,s})^2}{T^2}\right)}$,  
$t_{p}=t_{s}+\tau$ with $t_{p,s}$ - the centers of the pulses; $\Omega_{\text{max}}=2\pi~40$ MHz, full-width at half maximum duration $T_{\text{fwhm}} = 2\sqrt{\log{(2)}}T=0.2~\mu$s, and time delay $\tau=1.2 T$. 
The total duration is $3T_{\text{fwhm}}+\tau$. 
(c) The system is initially in $|1\rangle$ and the dark state $|d\rangle$. The ratio $\Omega_{p}(t)/\Omega_{s}(t)$ changes the mixing angle $\vartheta$ and $|d\rangle$ gets aligned with $-|3\rangle$ in the end. (d) When the system is initially in $|3\rangle$ it is also in a superposition of $|b_{\pm}\rangle$. 
As $\vartheta$ changes, so does the composition of $|b_{\pm}\rangle$.
(e) Adiabatic evolution causes population transfer from $|1\rangle$ to $|3\rangle$. (f) In contrast, the initial population of $|3\rangle$ transfers to $|1\rangle$ and/or $|2\rangle$, depending on the pulse area $\mathcal{A}=\int\Omega_{\text{rms}}(t)\text{d}t$ \cite{Vitanov2017}. 
Numerically simulated relative error $\epsilon =\left|1-\frac{w(t)}{\widetilde{w}(t)}\right|$ (see text for details) of the final polarization vs.\, $\Omega_{\text{max}}$ and pulse delay when the system is initially (g) in state $|1\rangle$ or (h) in state $|3\rangle$. 
	}
	\label{figSTIRAP}
\end{centering}
\end{figure}

 We consider a three-state, e.g., $\Lambda$-system as shown in Fig.\,\ref{figSTIRAP}. We aim for robust population exchange between the state $\ket{1}$ and state $\ket{3}$, mediated via couplings to an intermediate state $\ket{2}$ by two fields. We term the two fields pump (p) and Stokes (s), as is standard to the widely used STIRAP technique \cite{Vitanov2017}. 
 %

The coupling strengths are given by the Rabi frequencies $\rabi{p}(t)=-\mu_{12}\mathcal{E}_p(t)/\hbar$ and $\rabi{s}(t)=-\mu_{23}\mathcal{E}_s(t)/\hbar$ \cite{Shore2011}. Here, $\mu_{ij}$ are the transition dipole moments and $\mathcal{E}_{p/s}(t)$ are the time-varying envelopes of the electric fields. The system dynamics are described by the time-dependent Schr\"odinger equation $i\partial_t c(t)=H_\text{RWA}(t)c(t)$, where $c(t)=[c_1(t),c_2(t),c_3(t)]^T$ is a column vector with the probability amplitudes of the three states and the Hamiltonian in the rotating wave approximation is given by ($\hbar=1$) \cite{Vitanov2017,Shore2011,Shore1990}
\begin{align}
\label{eqHamiltonianRWA}
H_\text{RWA}(t)=
\frac{1}{2}
\begin{pmatrix}
-\delta & \rabi{p}(t) & 0 \\
\rabi{p}(t) & 2\Delta & \rabi{s}(t) \\
0 & \rabi{s}(t) & \delta
\end{pmatrix},
\end{align}
where the detunings of the driving fields from the corresponding resonances are defined as $\Delta_p=\omega_\text{12}-\omega_p$ and $\Delta_s=\omega_\text{32}-\omega_s$ with the single-photon detuning given by $\Delta=(\Delta_p+\Delta_s)/2$, and the two-photon detuning $\delta=\Delta_p-\Delta_s$. Our goal is efficient and robust population swapping between states $|1\rangle$ and $|3\rangle$ (for any coherent superposition or a mixed state) without ideally populating state $|2\rangle$ during the interaction as it typically decays fast. We assume further that the system is on two-photon resonance, i.e., $\delta=0$, which is necessary for STIRAP and is usually experimentally feasible.


%
%
%
Stimulated Raman adiabatic passage (STIRAP), reviewed in Refs.\,\cite{Bergmann_2019,Vitanov2017}, offers robust, lossless methods for population transfer from an initial state $|1\rangle$ to a target state $|3\rangle$. However, its performance is usually sensitive to the initial state of the system. In order to analyze its applicability for ground state swapping, we consider the adiabatic basis, defined by the instantaneous eigenstates \cite{Vitanov2017}
\begin{subequations}
	\label{eqAdiabStates}
	\begin{align}
	\ket{b_{+}}&=\sin\vartheta\sin\phi\ket{1}+\cos\vartheta\sin\phi\ket{3}+\cos\phi\ket{2},\\
	\ket{b_{-}}&=\sin\vartheta\cos\phi\ket{1}+\cos\vartheta\cos\phi\ket{3}-\sin\phi\ket{2},\\
	\label{eqDarkState}
	\ket{d}&=\cos\vartheta\ket{1}-\sin\vartheta\ket{3},
	\end{align}
\end{subequations}
where the two mixing angles are given by
\begin{subequations}
	\begin{align}
	\label{eqMixingAngleTheta}
	\vartheta(t)&=\arctan\frac{\rabi{p}(t)}{\rabi{s}(t)},\\
	\label{eqMixingAnglePhi}
	\phi(t)&=\frac{1}{2}\arctan\frac{\Omega_{\text{rms}}(t)}{\Delta}
	\end{align}
\end{subequations}
with the root mean square Rabi frequency ${\Omega_{\text{rms}}(t)=\sqrt{|\rabi{p}(t)|^2+|\rabi{s}(t)|^2}}$. The eigenenergies of the system are $\epsilon_{\pm}(t)=\frac{1}{2}(\Delta\pm\sqrt{\Delta^2+\Omega_{\text{rms}}(t)^2})$, $\epsilon_0=0$.

The transformation to the adiabatic basis uses the rotation matrix
\begin{align}\label{eqRmatrix}
R(t)&=
\left(
\begin{array}{ccc}
 \sin \vartheta (t) \sin \phi (t) & \cos \phi (t) & \cos \vartheta (t) \sin \phi (t) \\
 \cos \vartheta (t) & 0 & -\sin \vartheta (t) \\
 \sin \vartheta (t) \cos \phi (t) & -\sin \phi (t) & \cos \vartheta (t) \cos \phi (t) \\
\end{array}
\right),
\end{align}
which allows us to obtain the Hamiltonian in the adiabatic basis
\begin{align}\label{eqHamiltonianAdiab}
&H_{\text{adiab}}(t)=R(t) H_\text{RWA}(t)R^{\dagger}(t)-i R(t)\partial_{t}R^{\dagger}(t)\\
&=\left(
\begin{array}{ccc}
 \epsilon_{+}(t) & i \vartheta'(t) \sin \phi (t) & i \phi'(t) \\
 -i \vartheta '(t) \sin \phi (t) & 0 & -i \vartheta '(t) \cos \phi (t) \\
 -i \phi'(t) & i \vartheta '(t) \cos \phi (t) & \epsilon_{-}(t) \\
\end{array}
\right)\approx\left(
\begin{array}{ccc}
 \epsilon_{+}(t) & 0 & 0 \\
 0 & 0 & 0 \\
 0 & 0 & \epsilon_{-}(t) \\
\end{array}
\right),\notag
\end{align}
where we assumed adiabatic evolution in the last equality, i.e., $|\vartheta '(t)|\ll |\epsilon_{\pm}(t)|$ and  $|\phi'(t)|\ll |\epsilon_{\pm}(t)|$, so we could neglect the effect of the off-diagonal elements of the Hamiltonian.

The time evolution of the system in the adiabatic basis is described by the propagator $U_{\text{adiab}}(t,t_0)$, which connects the values of the probability amplitudes of the adiabatic states at the initial and final times $t_0$ and $t$: $c_a(t) = U_{\text{adiab}}(t,t_0)c_a(t_0)$. The propagator takes the form $U_{\text{adiab}}(t,t_0)=\mathcal{T}\exp\left(-i\int_{t_0}^{t}H_{\text{adiab}}(t') dt'\right)=\exp\left(-i\int_{t_0}^{t}H_{\text{adiab}}(t') dt'\right)$, where $\mathcal{T}$ is the time-ordering operator and we used for the last equality that the Hamiltonian in the adiabatic approximation commutes with itself at different times $[H_{\text{adiab}}(t_1),H_{\text{adiab}}(t_2)]=0$. 
Then, we can calculate the propagator, which characterizes the time evolution in the bare basis $c(t) = U(t,t_0)c(t_0)$, where
\begin{align}\label{eqPropSTIRAP}
&U(t,t_0)=R^{\dagger}(t)U_{\text{adiab}}(t,t_0)R(t_0)=\left(
\begin{array}{ccc}
 A \widetilde{s}_{t} \widetilde{s}_{t_0}+\widetilde{c}_{t} \widetilde{c}_{t_0} & B \widetilde{s}_{t} & A \widetilde{c}_{t_0} \widetilde{s}_{t}-\widetilde{c}_{t}
   \widetilde{s}_{t_0} \\
 B \widetilde{s}_{t_0} & \text{C} & B \widetilde{c}_{t_0} \\
 A \widetilde{c}_{t} \widetilde{s}_{t_0}-\widetilde{c}_{t_0} \widetilde{s}_{t} & B \widetilde{c}_{t} & A \widetilde{c}_{t} \widetilde{c}_{t_0}+\widetilde{s}_{t}
   \widetilde{s}_{t_0} \\
\end{array}
\right),
\end{align}
where we used the short-hand notation $\widetilde{c}_{t_0}=\cos{\vartheta(t_0)}$,
$\widetilde{s}_{t_0}=\sin{\vartheta(t_0)}$,
$\widetilde{c}_{t}=\cos{\vartheta(t)}$,
$\widetilde{s}_{t}=\sin{\vartheta(t)}$,
$A=e^{-i \eta _+} \sin ^2\phi+e^{-i \eta _-} \cos ^2\phi$,
$B=\frac{1}{2}\left(e^{-i \eta _+}-e^{-i \eta _-}\right) \sin 2\phi$,
$C=e^{-i \eta _-} \sin ^2\phi+e^{-i \eta _+} \cos ^2\phi$,
$\eta _{+}=\int_{t_0}^{t}\epsilon_{+}(t')dt'$,
$\eta _{-}=\int_{t_0}^{t}\epsilon_{-}(t')dt'$,
and we assumed that the mixing angle $\phi(t)=\phi(t_0)=\phi$ is the same at the beginning and at the end of the interaction, which is usually valid for STIRAP.

\section{Condition for two-state swapping \label{secSWAPcondition}}

We demonstrate in this section that performing efficient and robust population transfer from state $|1\rangle$ to $|3\rangle$ and vice versa is a sufficient condition for performing population swapping of arbitrary populations of the two states for any coherent interaction.
Speficically, if the system is initially in state $|1\rangle$ its state vector is $c(t_0)=[1,0,0]^{T}$. The state vector after the interaction is given by $c(t)=U(t,t_0)c(t_0)=[U_{1,1},U_{2,1},U_{3,1}]^{T}$, where $U_{i,j}$ is the $i,j$-th element of the propagator. Thus, successful population transfer from state $|1\rangle$ to $|3\rangle$ ensures  $|U_{1,1}|=|U_{2,1}|=0$ and $|U_{3,1}|=1$.
Similarly, successful population transfer from state $|3\rangle$ to $|1\rangle$ requires  $|U_{1,3}|=1$ and $|U_{2,3}|=|U_{3,3}|=0$. Finally, one can use that $U(t,t_0)U(t,t_0)^{\dagger}=I$, which requires $|U_{2,2}|=1$ and $|U_{1,2}|=|U_{3,2}|=0$. Thus, successful population transfer from state $|1\rangle$ to $|3\rangle$ and vice versa ensures
\begin{align}\label{Eq:swap_propagator}
U(t,t_0)&=
\left(
\begin{array}{ccc}
 0 & 0 & \e^{i\alpha_3} \\
 0 & \e^{i\alpha_2} & 0 \\
 \e^{i\alpha_1} & 0 & 0 \\
\end{array}
\right),
\end{align}
where $\alpha_{k},~k=1,...,3$ are phases, which depend on the specific interaction.
If the initial state of the system is an arbitrary coherent superposition of the three states $c(t_0)=[c_1(t_0),c_2(t_0),c_3(t_0)]^{T}$, the final state after the interaction is $c(t
)=U(t,t_0)c(t_0)=\left[e^{i \alpha_3 } c_3(t_0),e^{i \alpha_2 } c_2(t_0),e^{i \alpha_1 } c_1(t_0)\right]^{T}$, i. e., the populations of the states $|1\rangle$ and $|3\rangle$: $P_1(t)=|e^{i\alpha_3} c_3(t_0)|^2=P_3(t_0)$ and $P_3(t)=|e^{i\alpha_1} c_1(t_0)|^2=P_1(t_0)$ are swapped.
It is also straightforward to show that the populations of states $|1\rangle$ and $|3\rangle$ are interchanged if the system is initially in a mixed state, so it is characterized by its density matrix. Specifically, when the system is initially in a (partially) mixed state involving $|1\rangle$ and $|3\rangle$, its density matrix takes the form
\begin{align}
\rho(t_0)=
\left(
\begin{array}{ccc}
 P_1(t_0) & 0 & e^{-\Gamma}e^{i\chi} \sqrt{P_1(t_0) P_3(t_0)} \\
 0 & P_2 & 0 \\
e^{-\Gamma}e^{-i\chi}\sqrt{P_1(t_0) P_3(t_0)} & 0 & P_3(t_0) \\
\end{array}
\right),
\end{align}
where the diagonal element $\rho_{k,k}(t_0)=P_k(t_0)$ is equal to the population of state $|k\rangle,k=1,...,3$, while $\rho_{1,3}(t_0)=\rho_{3,1}(t_0)^{\ast}=e^{-\Gamma}e^{i \chi}\sqrt{P_1(t_0) P_3(t_0)}$ is the coherence, which characterizes the (partially incoherent) superposition between states $|1\rangle$ and $|3\rangle$, where $\chi$ is its phase and the parameter $e^{-\Gamma}$ characterizes the degree of decoherence ($\Gamma\ge 0$). When $\Gamma\to 0$, $e^{-\Gamma}\to 1$ and the system is in a pure coherent superposition state. On the contrary, when $\Gamma\to +\infty$, $e^{-\Gamma}\to 0$ and the system is in a fully mixed, or incoherent, state. We assumed for simplicity of presentation and without loss of generality that all coherences, involving state $|2\rangle$ have completely decayed.
The density matrix after the interaction is given by
\begin{align}
\rho(t)&=U(t,t_0)\rho(t_0)U^{\dagger}(t,t_0)\notag\\
&=\left(
\begin{array}{ccc}
 P_3(t_0) & 0 & e^{-\Gamma}e^{-i(\chi +\alpha _1-\alpha _3)} \sqrt{P_1(t_0) P_3(t_0)} \\
 0 & P_2 & 0 \\
 e^{-\Gamma}e^{i(\chi +\alpha _1-\alpha _3)} \sqrt{P_1(t_0) P_3(t_0)} & 0 & P_1(t_0) \\
\end{array}
\right),
\end{align}
and the populations of the two states $P_1(t)=\rho_{1,1}(t)=P_3(t_0)$ and $P_3(t)=\rho_{3,3}(t)=P_1(t_0)$ are interchanged. Thus, successful population transfer from state $|1\rangle$ to $|3\rangle$ and vice versa is a sufficient condition for swapping of arbitrary populations of states $|1\rangle$ and $|3\rangle$, including for coherent superpositions or mixed states. Next, we analyze several versions of STIRAP and their performance for population swapping by first characterizing their efficiency for population transfer from state $|1\rangle$ to $|3\rangle$ and vice versa. If a technique is successful in the transfers from both of them, then it should also perform successful swapping, as demonstrated above. In order to confirm this we also explicitly simulate numerically the efficiency of population swapping for arbitrarily chosen fully mixed and coherent superposition states for each version of STIRAP.

\begin{figure}[t]
\begin{centering}
	\includegraphics[width=0.82\columnwidth]{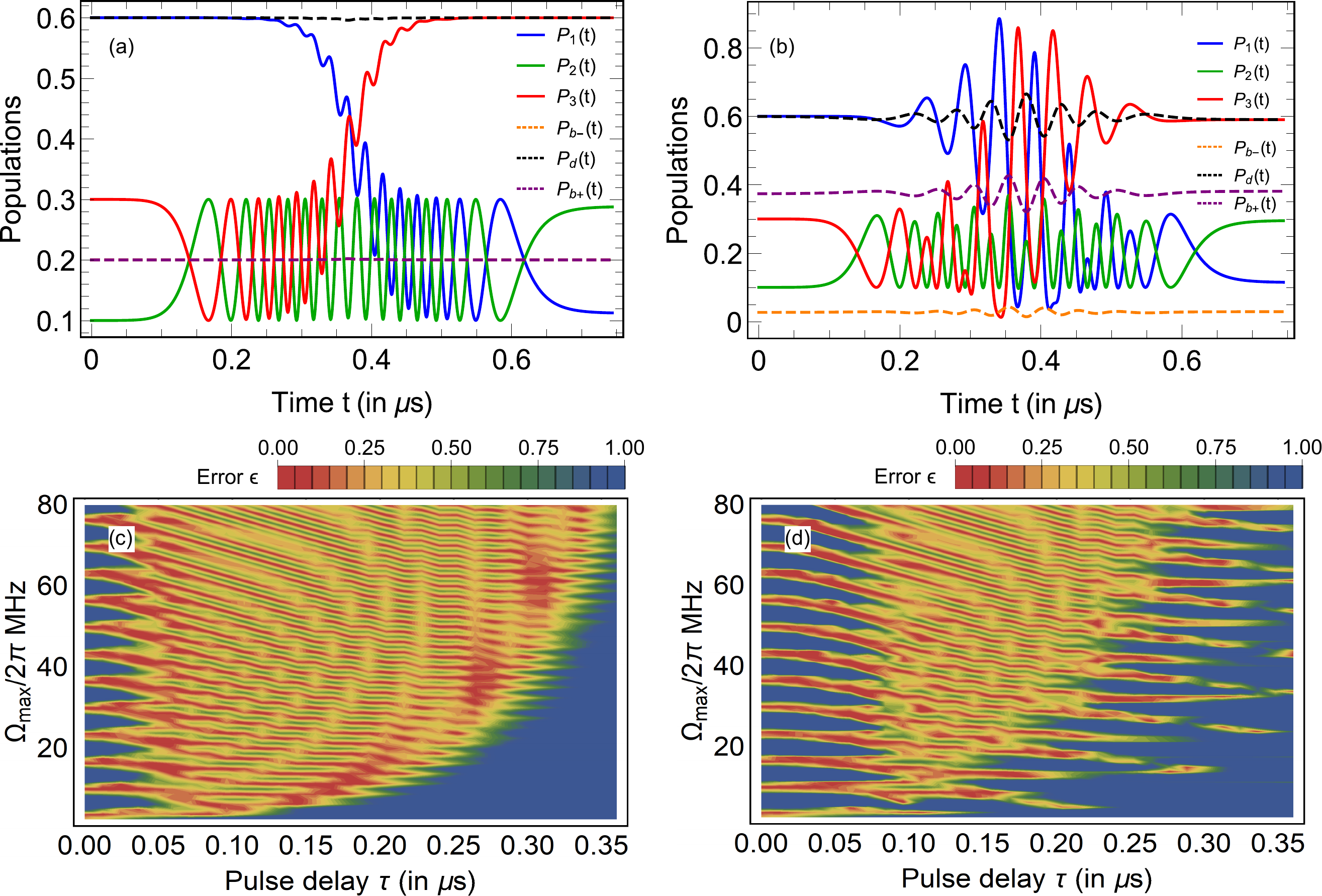}
\caption{Numerically simulated time evolution of the populations for resonant STIRAP with all pulse parameters the same as in Fig. \ref{figSTIRAP} for
(a) an initial fully mixed state with populations $P_1(t_0)=0.6$, $P_2(t_0)=0.1$, $P_3(t_0)=0.3$, (b) an initial pure state with probability amplitudes $c(t_0)=\left[\sqrt{0.6},\sqrt{0.1},\sqrt{0.3}\right]^{T}$.  
In both cases the population of state $|1\rangle$ is transferred to state $|3\rangle$ but the transfer efficiency of the population of state $|3\rangle$ to $|1\rangle$ is highly sensitive to the pulse area $\mathcal{A}$.
Numerically simulated magnitude of the relative error of the final polarization $\epsilon$ vs. pulse delay and the maximum Rabi frequency of the pump and Stokes pulses
(c) for an initial fully mixed state with populations as in (a), and (d) for an initial pure state as in (b). The results show a highly oscillatory behavior due to dephasing in the adiabatic basis that makes resonant STIRAP unsuitable for population swapping of arbitrary states.
	}
	\label{fig_res_STIRAP_partial}
\end{centering}
\end{figure}

\section{Resonant STIRAP\label{secResSTIRAP}}

We first consider the tradtional, resonant case of STIRAP, where the single-photon detuning $\Delta=0$ \cite{Vitanov2017}. STIRAP requires pump and Stokes pulses in the so-called counter-intuitive order \footnote{We note that after understanding the principles of STIRAP this ``counter-intuitive'' pulse order, in fact, becomes perfectly intuitive.}, as shown in Fig.\,\ref{figSTIRAP}, where the Stokes pulse precedes the pump pulse by a time delay $\tau$.
It is well known that if the system is initially prepared in state $\ket{1}$, STIRAP transfers the population completely  to state $\ket{3}$ via the dark state $\ket{d}$, ideally, without populating state $\ket{2}$ (see Fig.\,\ref{figSTIRAP}(c), which shows the composition of the dark state and Fig.\,\ref{figSTIRAP}(d) for the respective population evolution).
The dynamics can be understood by analyzing the evolution in the mixing angle $\vartheta$. As the latter changes from $\vartheta=0$ to $\vartheta=\pi/2$, the dark state evolves from $\ket{d}=\ket{1}$ to $\ket{d}=-\ket{3}$.
During the process, we must maintain adiabaticity, i.e., the system must remain in the dark state at all times. This requires $\Omega_{\text{rms}}(t)\gg |\dot{\vartheta}|$.
For smooth pulses (e.g., with Gaussian temporal shape), this adiabatic condition transforms to the simpler form $\mathcal{A}=\int\Omega_{\text{rms}}(t)\text{d}t \gg 1$, i.e., the pulse area $\mathcal{A}$ has to be sufficiently large \cite{Vitanov2017}.
The larger $\mathcal{A}$, the closer the transfer efficiency approaches unity.

It is important to note that the population transfer efficiency of STIRAP depends strongly on the initial state and it is thus not suitable for inverting an arbitrary quantum state. Specifically, assuming the standard ``counterintuitive'' order of the pulses, where the Stokes pulse precedes the pump pulse, the population transfer is done via the dark state $|d\rangle$ if the system is initially in state $|1\rangle$. On the contrary, if the system is initially in state $|3\rangle$, the initial state in the adiabatic basis is an equal superposition of the bright states $|b_\pm\rangle$ when $\Delta=0$ (see Fig.\,\ref{figSTIRAP}(e), which shows the composition of the bright states and Fig.\,\ref{figSTIRAP}(f) for the respective population evolution). As a result, the system experiences phase evolution in the adiabatic basis during the interaction and the populations of the final states depend strongly on the overall accumulated phase, which is equal to the effective pulse area $\mathcal{A}$ \cite{Vitanov2017}
\begin{equation}
P_1=\cos^2{(\mathcal{A}/2)},~P_2=\sin^2{(\mathcal{A}/2)},~P_3=0.
\end{equation}
This also evident from the propagator of resonant STIRAP, which can be obtained from Eq. \eqref{eqPropSTIRAP} by taking $\phi(t_0)=\phi(t)=\phi=\pi/4$, $\vartheta(t_0)=0$, $\vartheta(t)=\pi/2$, $\epsilon_{+}=-\epsilon_{-}=\rabi{\text{rms}}/2$, so $\eta_{+}=-\eta_{-}=\mathcal{A}/2$
\begin{equation}
U_{\text{res.}}(t,t_0)=\left(
\begin{array}{ccc}
 0 & -i \sin \left(\mathcal{A}/2\right) & \cos \left(\mathcal{A}/2\right) \\
 0 & \cos \left(\mathcal{A}/2\right) & -i \sin \left(\mathcal{A}/2\right) \\
 -1 & 0 & 0 \\
\end{array}
\right).
\end{equation}
We note that the case when the system is initially in state $|3\rangle$ and we apply standard resonant STIRAP is equivalent to starting in state $|1\rangle$ and applying the opposite pulse ordering, where the pump pulse precedes the Stokes pulse. It is known that the population transfer efficiency of the latter is highly sensitive to experimental parameters \cite{Vitanov2017}.
The numerical simulations in Figs.\,\ref{figSTIRAP}(g) and \ref{figSTIRAP}(h) confirm the large difference in the dependence of the state swapping efficiency on the peak Rabi frequency and pulse delay when we start in states $|1\rangle$ and $|3\rangle$.
We used as a figure of merit in all 2D simulations the magnitude of the relative error $\epsilon$ of the final polarization $w(t)=P_1(t)-P_3(t)$ 
with respect to the final target polarization. The latter is $\widetilde{w}(t)=-\left(P_1(t_0)-P_3(t_0)\right)$, i.e., the inverse of the initial polarization as our goal is to swap the populations of the initial states and, thus,
$\epsilon =\left|1-\frac{w(t)}{\widetilde{w}(t)}\right|$. For example, when the system is initially in $|1\rangle$ or $|3\rangle$, $\epsilon=1-|w(t)|$, where we assumed that $w(t)$ and $\widetilde{w}(t)$ have the same sign. 
The numerical simulations confirm that state swapping is efficient and robust only when the system is initially in state $|1\rangle$ (see Fig. \ref{figSTIRAP}(g)) but the swap efficiency suffers highly oscillatory behavior when initially in $|3\rangle$ (see Fig. \ref{figSTIRAP}(h)). Thus, STIRAP efficiency and robustness depend strongly on the initial state and cannot be used for population swapping.

This is also confirmed in Fig.\,\ref{fig_res_STIRAP_partial} where we simulated the population dynamics and the error in the final polarization with the same pulse parameters as in Fig. \ref{figSTIRAP}. Figure \,\ref{fig_res_STIRAP_partial}(a) shows the evolution for
a mixed initial state with populations $P_1(t_0)=0.6$, $P_2(t_0)=0.1$, and $P_3(t_0)=0.3$, while Figure \,\ref{fig_res_STIRAP_partial}(b) demonstrates the population dynamics for a pure state, which is a coherent superposition of all three states with the same populations as in Fig. \,\ref{fig_res_STIRAP_partial}(a) and probability amplitudes $c(t_0)=\left[\sqrt{0.6},\sqrt{0.1},\sqrt{0.3}\right]^{T}$. Figures \,\ref{fig_res_STIRAP_partial}(c) and \,\ref{fig_res_STIRAP_partial}(d) show the high sensitivity of the swap efficiency of resonant STIRAP to the experimental parameters for the two cases, as expected from theory.

\begin{figure}[t]
\begin{centering}
	\includegraphics[width=0.82\columnwidth]{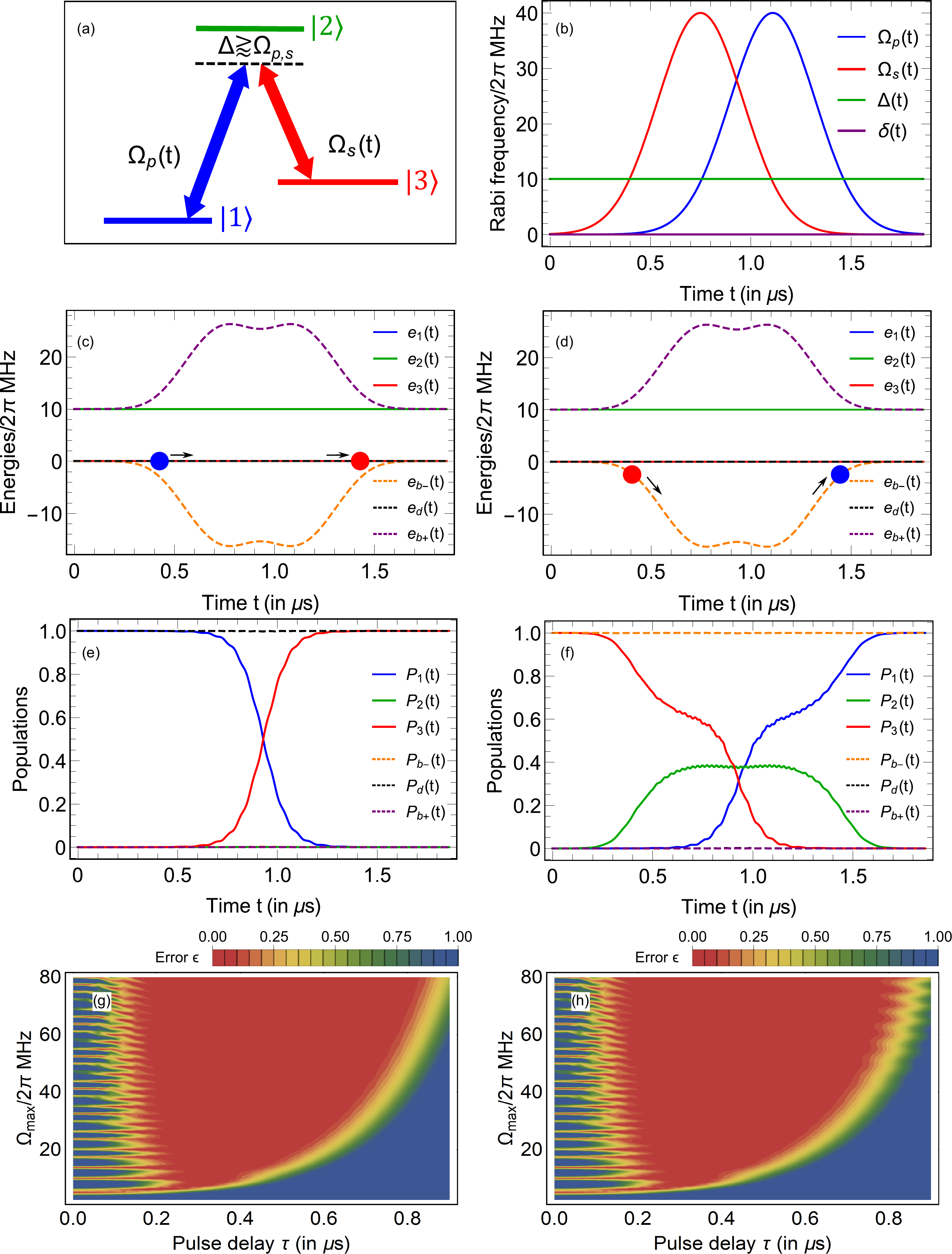}
\caption{(a) Coupling scheme for STIRAP with a moderate detuning, (b) the respective pulse sequence with the same parameters as in Fig.\,\ref{figSTIRAP}, except for  $\Delta=2\pi~10$ MHz and the longer $T_{\text{fwhm}} = 2\sqrt{\log{(2)}}T=0.5\,\mu$s to maintain adiabaticity. (c) The system is initially in state $|1\rangle$ and the dark state $|d\rangle$. As the mixing angle $\vartheta$ changes $|d\rangle$ aligns with $-|3\rangle$ in the end. 
(d) The system is initially in $|3\rangle$ and $|b_{-}\rangle$. As $|b_{-}\rangle$ is an eigenstate, the system is not affected by dephasing in the adiabatic basis, in contrast to resonant STIRAP. 
(e) Population transfer from $|1\rangle$ to $|3\rangle$ via the dark state $|d\rangle$. 
(f) Population transfer from $|3\rangle$ to $|1\rangle$ via the bright state $|b_{-}\rangle$ is efficient and robust, in contrast to resonant STIRAP. The main drawback is the high intermediate population in state $|2\rangle$. Numerically simulated relative error $\epsilon$ of the final polarization vs.\, $\Omega_{\text{max}}$ and pulse delay when the system is initially (g) in state $|1\rangle$, (h) or in state $|3\rangle$.
	}
	\label{figmSTIRAP}
\end{centering}
\end{figure}

\begin{figure}[t]
\begin{centering}
	\includegraphics[width=0.82\columnwidth]{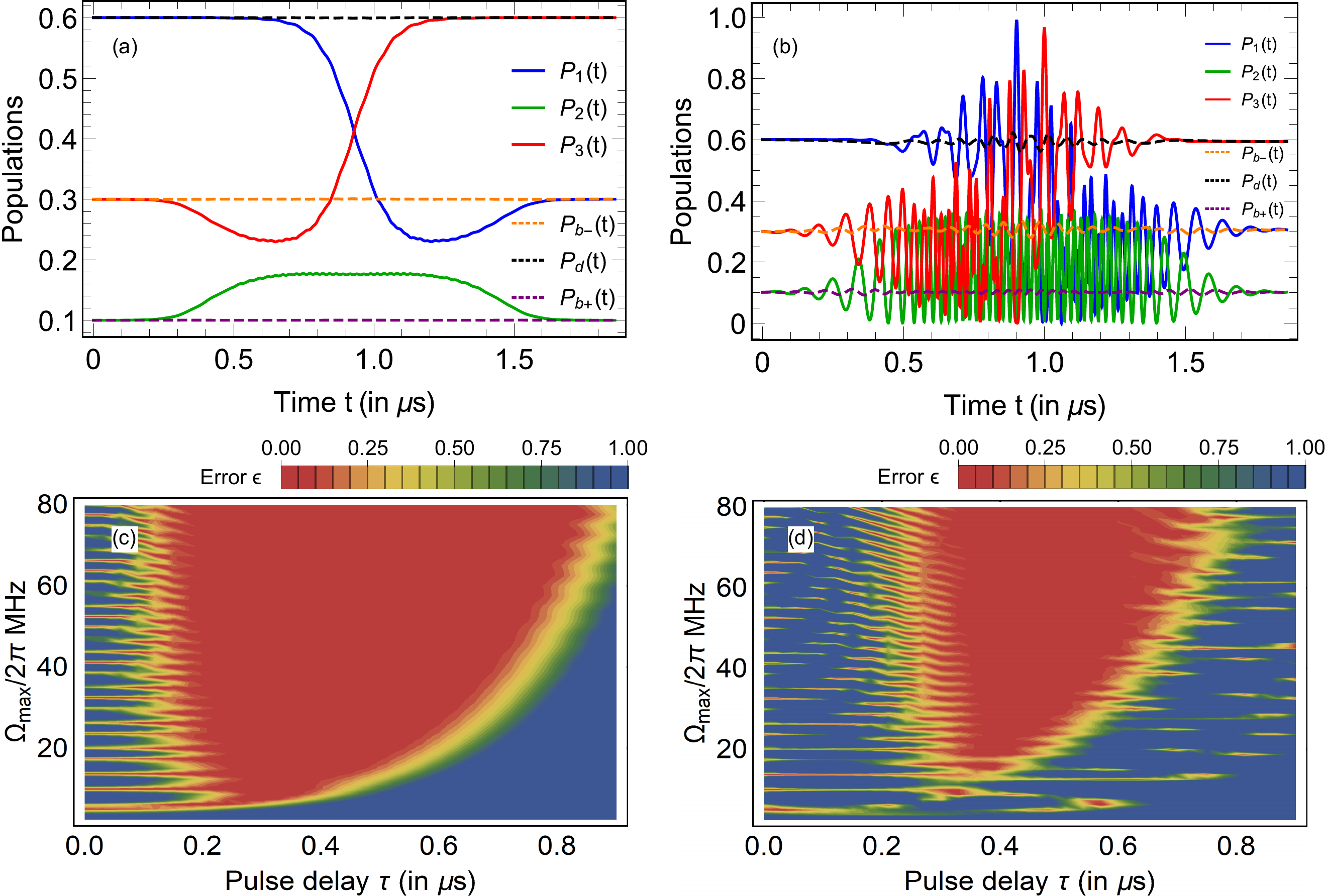}
\caption{Numerically simulated evolution of populations for STIRAP with a moderate detuning with all pulse parameters the same as in Fig. \ref{figmSTIRAP} for
(a) an initial fully mixed state with populations $P_1(t_0)=0.6$, $P_2(t_0)=0.1$, $P_3(t_0)=0.3$, and (b) an initial pure state with probability amplitudes $c(t_0)=\left[\sqrt{0.6},\sqrt{0.1},\sqrt{0.3}\right]^{T}$.  
Numerically simulated relative error of the final polarization $\epsilon =\left|1-\frac{w(t)}{\widetilde{w}(t)}\right|$ (c) for an initial fully mixed state as in (a), and (d) for an initial pure state coherent superposition state as in (b). It is evident that STIRAP with a moderate detuning performs efficient and robust state swapping for an arbitrary initial state. The slightly narrower range of experimental parameters with robust performance for a pure state is due to  coherences that dephase in the adiabatic basis and make the process more sensitive to the fulfillment of the adiabatic condition.
	}
	\label{figmSTIRAP_partial}
\end{centering}
\end{figure}

\section{STIRAP with a moderate detuning \label{secModDetunedSTIRAP}}

We now consider the case of STIRAP in the presence of nonzero single-photon detuning, e.g., when $\Delta\gtrapprox 0.1\rabi{\text{max}}$ and the system is in two-photon resonance ($\delta=0$).
We have labelled it STIRAP with a moderate detuning to emphasize that it is an intermediate case between resonant STIRAP and the one with a large detuning where we can typically adiabatically eliminate the intermediate state.
However, we note that the required frequency offset could actually be quite small, i.e., a fraction of the peak Rabi frequency of the two fields. The necessary single-photon detuning is determined by the requirement to lift the degeneracy of the bright states $|b_{\pm}\rangle$ and the adiabatic approximation, so that there is no population transfer between the adiabatic states during the interaction.
%
We assume that the evolution of the system is adiabatic and we can analyze it using the adiabatic states in Eqs.\,\eqref{eqAdiabStates}, similarly to the theoretical analysis of STIRAP with an intermediate-level detuning in \cite{Vitanov1997}.
When the system is initially in state $|1\rangle$, the adiabatic population transfer takes place via the dark state $\ket{d}$, similarly to the resonant case (see Fig.\,\ref{figmSTIRAP}(c)). In contrast, when the system is initially in state $|3\rangle$, adiabatic population transfer goes via the bright state $\ket{b_{-}}$ (or $\ket{b_{+}}$), assuming $\Delta>0$ (or $\Delta<0$), as the mixing angle $\phi\to 0$ (or $\phi\to \pi/2$) \cite{Klein2008}. In the following analysis, we assume $\Delta>0$ for simplicity of presentation and without loss of generality, so the system is initially in the bright state $\ket{b_{-}}$ (see Fig.\,\ref{figmSTIRAP}(f)). As the system is now in one of the eigenstates in the adiabatic basis, the robustness of population transfer is not affected by phase evolution in this basis in contrast to resonant STIRAP.
We note that this case is equivalent to being initially in state $|1\rangle$ and applying STIRAP with an ``intuitive'' ordering, where the pump pulse precedes the Stokes pulse, also known as bright STIRAP, as it takes place via one of the bright states \cite{Vitanov2017,Klein2007}.
Thus, STIRAP with a moderate detuning allows for a smooth adiabatic population inversion when the system is initially in state $|1\rangle$ and in state $|3\rangle$. The numerical simulations in Figs.\,\ref{figmSTIRAP}(g) and \ref{figmSTIRAP}(h) confirm the high robustness of the process when starting in both $|1\rangle$ and $|3\rangle$, assuming negligible decay from $|2\rangle$. Thus, based on the analysis in section \ref{secSystem}, we conclude that STIRAP with a moderate detuning can perform efficient and robust population swapping of states $|1\rangle$ and $|3\rangle$ for arbitrary initial states. The main difference when starting in $|3\rangle$ is the high intermediate population of state $|2\rangle$, which can generally decay to the other states and lead to a loss of fidelity.

The ability to perform successful population swapping is also confirmed by analyzing the propagator of STIRAP with a non-zero single-photon detuning. We can obtain the latter from Eq. \eqref{eqPropSTIRAP} by taking $\phi=0$, $\vartheta(t_0)=0$, $\vartheta(t)=\pi/2$ and assuming $\Delta>0$ without loss of generality
\begin{equation}\label{eqPropDetSTIRAP}
U_{\text{det.}}(t,t_0)=\left(
\begin{array}{ccc}
 0 & 0 & e^{-i \eta _-} \\
 0 & e^{-i \eta _+} & 0 \\
 -1 & 0 & 0 \\
\end{array}
\right).
\end{equation}
It is evident by comparison with Eq. \eqref{Eq:swap_propagator} that STIRAP with a moderate detuning can perform successful swapping of the populations of states $|1\rangle$ and $|3\rangle$.

Finally, we perform numerical simulations for population swapping for arbitrary mixed and coherent superposition pure states in Fig.\,\ref{figmSTIRAP_partial} with the same pulse parameters as in Fig. \ref{figmSTIRAP}. Figure \,\ref{figmSTIRAP_partial}(a) shows the evolution for a mixed initial state with populations $P_1(t_0)=0.6$, $P_2(t_0)=0.1$, and $P_3(t_0)=0.3$. In addition, Figure \,\ref{figmSTIRAP_partial}(b) describes the corresponding population dynamics for a pure state, which is a coherent superposition of all three states with the same populations as in Fig. \ref{figmSTIRAP_partial}(a) and probability amplitudes $c(t_0)=\left[\sqrt{0.6},\sqrt{0.1},\sqrt{0.3}\right]^{T}$. Both Figs. \,\ref{figmSTIRAP_partial}(c) and \,\ref{figmSTIRAP_partial}(d) show that the swap efficiency is high and robust to the experimental parameters for both the mixed and pure coherent superposition states, as expected from theory. The slightly narrower range of experimental parameters with robust performance for a pure state is due to the presence of coherences that dephase in the adiabatic basis, which makes the process a bit more sensitive to the fulfillment of the adiabatic condition.
In conclusion, STIRAP with a moderate detuning can be used for efficient and robust swapping of the ground state populations of a qubit in an unknown state, as long as the evolution of the system is adiabatic and there is negligible decay from the intermediate state $|2\rangle$ during the interaction.




\begin{figure}[t]
\begin{centering}
	\includegraphics[width=0.8\columnwidth]{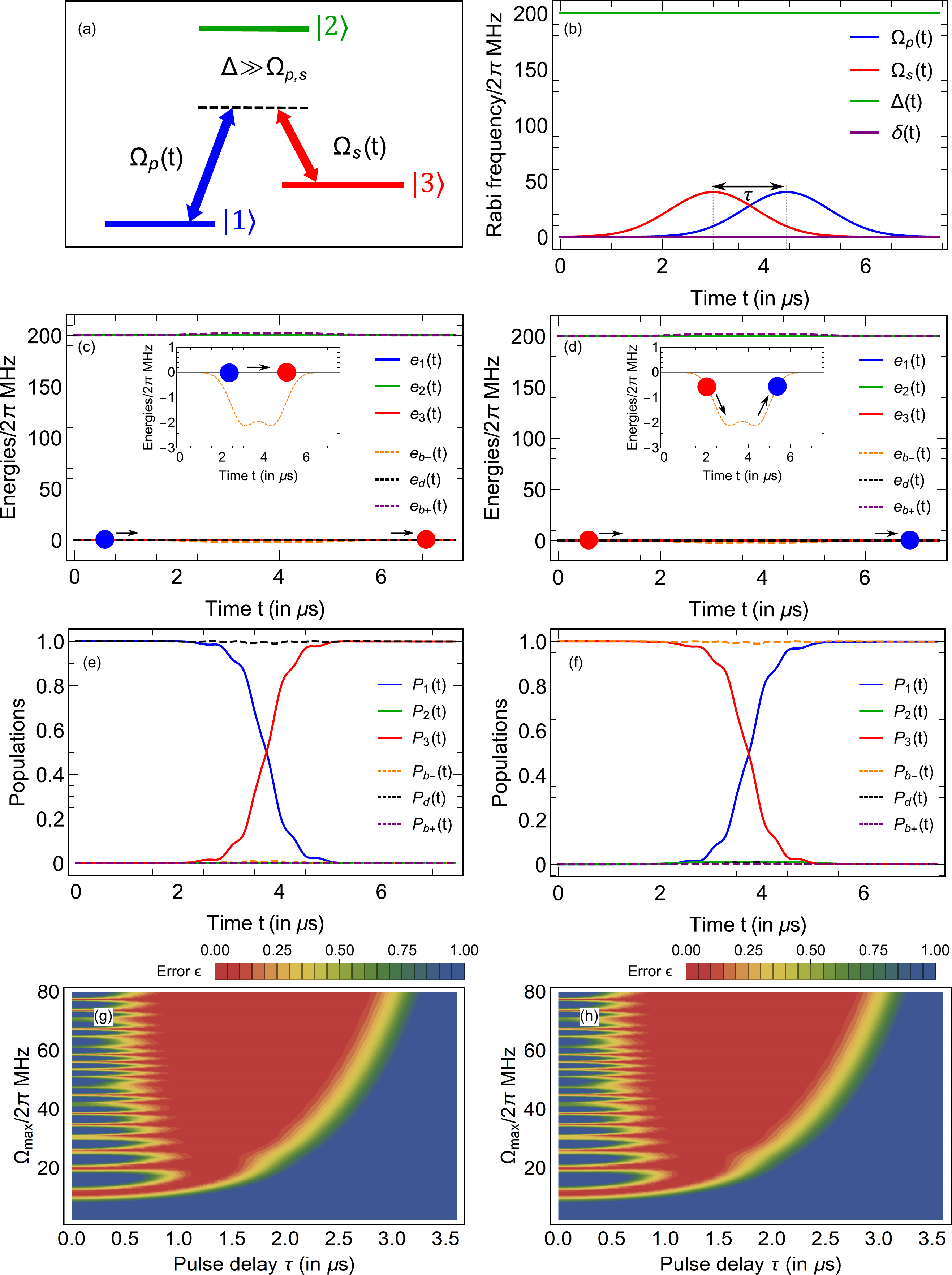}
\caption{(a) Coupling scheme for STIRAP with a large detuning, (b) the respective pulse sequence with the same parameters as in Fig.\,\ref{figSTIRAP}, except for  $\Delta=2\pi~200$ MHz and the longer $T_{\text{fwhm}} = 2\sqrt{\log{(2)}}T=2\,\mu$s to maintain adiabaticity. (c) The system is initially in state $|1\rangle$ and the dark state $|d\rangle$. As $\vartheta$ changes $|d\rangle$ aligns with $-|3\rangle$ in the end. 
(d) The system is initially in $|3\rangle$ and $|b_{-}\rangle$. As $|b_{-}\rangle$ is an eigenstate, the system is not affected by dephasing in the adiabatic basis. 
(e) Population transfer from $|1\rangle$ to $|3\rangle$ via the dark state $|d\rangle$. 
(f) Population transfer from $|3\rangle$ to $|1\rangle$ via the bright state $|b_{-}\rangle$ is efficient and robust, similarly to STIRAP with a moderate detuning. The main advantage is the negligible intermediate population in state $|2\rangle$. Numerically simulated relative error $\epsilon$ of the final polarization vs.\, $\Omega_{\text{max}}$ and pulse delay when the system is initially in (g) state $|1\rangle$ (h) or state $|3\rangle$.
	}
	\label{figdSTIRAP}
\end{centering}
\end{figure}

\begin{figure}[t]
\begin{centering}
	\includegraphics[width=0.82\columnwidth]{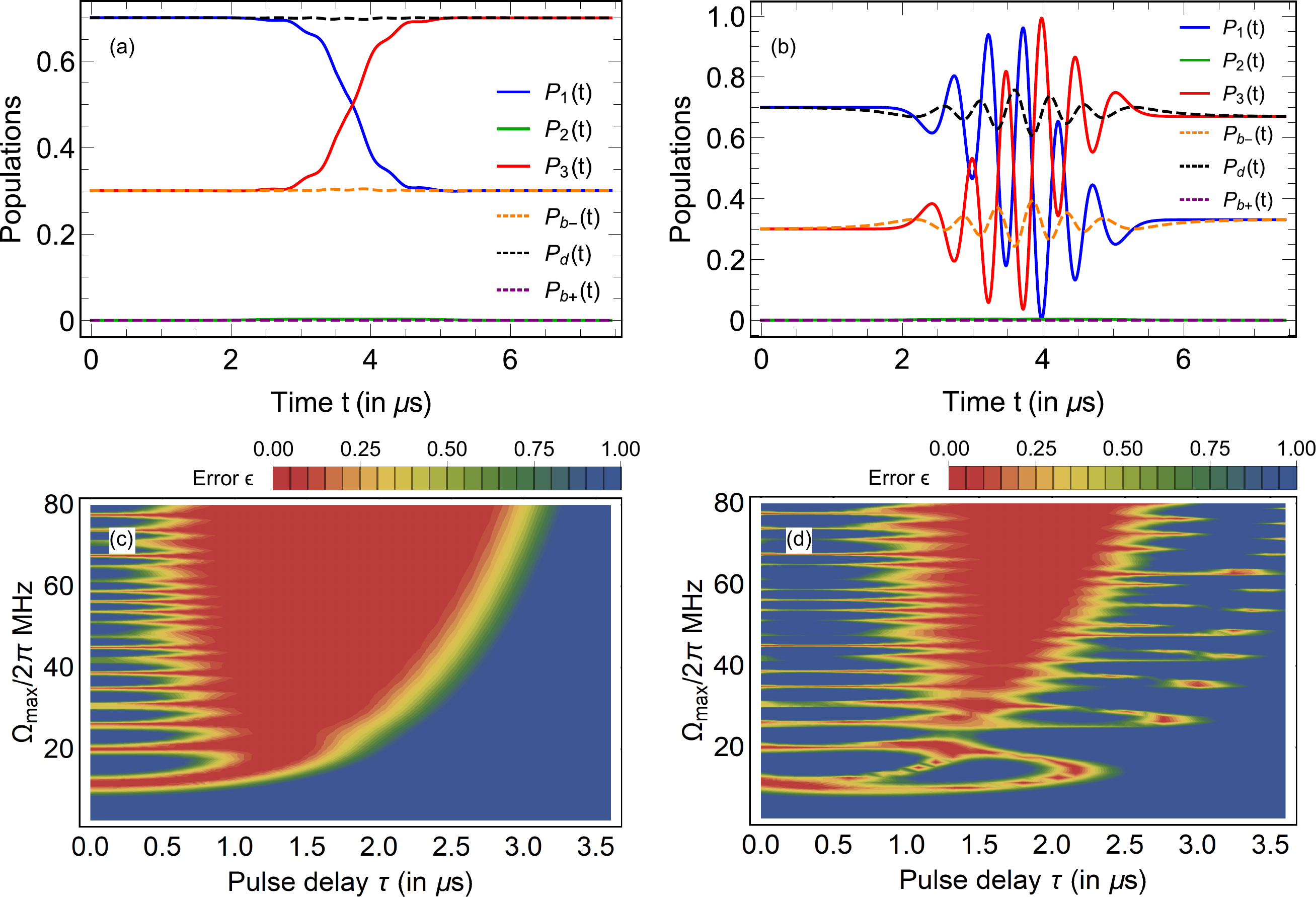}
\caption{Numerically simulated evolution of populations for STIRAP with a large detuning with all pulse parameters the same as in Fig. \ref{figdSTIRAP} for
(a) an initial fully mixed state with populations $P_1(t_0)=0.7$, $P_2(t_0)=0$, $P_3(t_0)=0.3$, and (b) an initial pure state with probability amplitudes $c(t_0)=\left[\sqrt{0.7},0,\sqrt{0.3}\right]^{T}$.  
(f) Numerically simulated relative error of the final polarization $\epsilon =\left|1-\frac{w(t)}{\widetilde{w}(t)}\right|$ for
(c) an initial fully mixed state as in (a) and (d) an initial pure state coherent superposition state as in (b). It is evident that STIRAP with a large detuning performs efficient and robust state swapping for an arbitrary initial state with negligible intermediate change in the population of the intermediate state $|2\rangle$. The slightly narrower range of experimental parameters with robust performance for a pure state is due to  coherences that dephase in the adiabatic basis and make the process more sensitive to the fulfillement of the adiabatic condition, similarly to the moderately detuned case.
	}
	\label{figdSTIRAP_partial}
\end{centering}
\end{figure}

\section{STIRAP with a large detuning \label{secHighDetunedSTIRAP}}

We now consider the case of STIRAP with a large single-photon detuning ${\Delta\gg\rabi{\text{max}}}$. It is known that we can, in principle, adiabatically eliminate the excited state $\ket{2}$ from the system for the theoretical description.
This transforms the three-level system into an effective two-level scheme and allows for state-independent state inversion \cite{Vitanov1997}.
%
Nevertheless, we consider the evolution of the system without applying adiabatic elimination in order to expand the applicability of the analysis and compare directly with resonant STIRAP and STIRAP with a moderate detuning. A detailed description of the adiabatic elimination  approach and a comparison to the following analysis is provided in \ref{secAdiabElimination}.

When the system is initially in state $|1\rangle$, the adiabatic population transfer takes place via the dark state $\ket{d}$, similarly to the resonant case (see Fig.\,\ref{figdSTIRAP}(c)). In contrast, when the system is initially in state $|3\rangle$, adiabatic population transfer goes via the bright state $\ket{b_{-}}$ (assuming $\Delta>0$), as in the case of moderate detuning \cite{Klein2008} (see Fig.\,\ref{figdSTIRAP}(d)). As the system is in one of the eigenstates in the adiabatic basis, the robustness of population transfer is not affected by phase evolution in this basis, in contrast to resonant STIRAP. Thus, STIRAP with a large detuning allows for smooth, robust, adiabatic transfer when the system is initially in state $|3\rangle$.

In contrast to the moderate-detuning case, there is negligible population in the intermediate state $|2\rangle$, $P_2(t)\sim \rabi{\text{rms}}(t)^2/\Delta^2$, during the process (see Fig.\,\ref{figdSTIRAP}(e) and Fig.\,\ref{figdSTIRAP}(f)), which makes STIRAP with a large detuning robust to fast decay from the intermediate state $|2\rangle$.
The trade-off is the longer interaction time needed to maintain adiabaticity. Specifically, the adiabatic states $\ket{d}$ and $\ket{b_{-}}$ have the smallest energy difference (assuming $\Delta>0$), so the adiabatic condition becomes $|\vartheta '(t)|\ll |\epsilon_{-}(t)-\epsilon_{d}(t)|=|\epsilon_{-}(t)|$. In the limit $\Omega _{\text{rms}}(t)\ll \Delta$ we obtain $\epsilon_{-}\approx-\frac{\Omega _{\text{rms}}(t){}^2}{4 \Delta }$, which scales as $\Delta^{-1}$ while $\vartheta '(t)$ does not depend on $\Delta$. Thus, the duration of the process should increase linearly with $\Delta$ in order to maintain adiabaticity for large detuning. The intermediate population $P_2(t)\sim\rabi{\text{rms}}(t)^2/\Delta^2$ scales as an inverse quadratic function with $\Delta$, so there is a range of large detunings where STIRAP can minimize population loss from state $|2\rangle$ and keep a reasonable total duration.

The numerical simulations in Figs.\,\ref{figdSTIRAP}(g) and \ref{figdSTIRAP}(h) confirm the high efficiency and robustness of the process, which is independent from the initial state, i.e., $|1\rangle$ or $|3\rangle$.
Assuming adiabatic evolution, the propagator of STIRAP with a large detuning at the end of the interaction is given in Eq. \eqref{eqPropDetSTIRAP}. 
It also confirms that it can perform successful swapping of the populations of states $|1\rangle$ and $|3\rangle$.

Finally, we perform numerical simulations for population swapping for arbitrary mixed and coherent superposition pure states in Fig.\,\ref{figdSTIRAP_partial}. In all cases, the swap efficiency is high and robust to the experimental parameters, as expected from theory. The slightly narrower range of experimental parameters with robust performance for a pure coherent superposition state is again due to the presence of coherences that dephase in the adiabatic basis, similarly to STIRAP with a moderate detuning.
All simulations confirm that STIRAP with a large detuning is very suitable for efficient and robust swap of the populations of a qubit in an arbitrary, unknown state. As the population of the intermediate state $|2\rangle$ does not change during the interaction, the technique is especially useful when the intermediate state is fast-decaying.

\section{Discussion\label{secDiscussion}}

Our analysis show that efficient and robust state-independent state swapping is possible with STIRAP unlike the widely accepted notion that its performance is sensitive to the initial state of the system. We only require a non-zero single-photon detuning, while maintaining adiabaticity. We have labeled this case STIRAP with a moderate detuning for simplicity of presentation but the frequency offset can actually be quite small, e.g., of the order of a fraction of the peak Rabi frequency of the pump or Stokes fields. The single-photon detuning lifts the degeneracy of the bright adiabatic states, aligns each bare state with an eigenstate in the adiabatic basis at the beginning of the interaction, and thus reduces the effect of dephasing in the adiabatic basis, making efficient population swapping possible.
STIRAP with a moderate detuning is faster than the highly detuned case as the effective two-photon couplings are stronger, so maintaing adicabaticity requires shorter total interaction time. 
However, it allows for some intermediate population in state $|2\rangle$, so it is preferable when the latter is not fast decaying. A feasible example is double-quantum manipulation of the $|\pm 1\rangle$ subsystem of the ground state of NV centers. This qubit in principle allows for faster phase accumulation and more efficient quantum sensing than using, for example, the ground states $|0\rangle$ and $|-1\rangle$ \cite{Degen2017RMP}.
When the intermediate state is fast decaying, population swapping requires STIRAP with a large detuning, so that state $|2\rangle$ is not populated in the process for any initial state. The main drawback is the longer duration of the process, which is necessary in order to maintain adiabaticity.

We note that STIRAP with a non-zero single-photon detuning can also be applicable for spin refocusing, e.g., in dynamical docoupling. This can be seen by considering the propagator of two identical STIRAP processes in the adiabatic approximation, which takes the form
\begin{equation}\label{eqProp2DetSTIRAP}
U_{\text{det.}}(2t_{\text{st}},t_{\text{st}}).U_{\text{det.}}(t_{\text{st}},0)=U_{\text{det.}}(t_{\text{st}},0).U_{\text{det.}}(t_{\text{st}},0)=\left(
\begin{array}{ccc}
 -e^{-i \eta _-} & 0 & 0 \\
 0 & e^{-2 i \eta _+} & 0 \\
 0 & 0 & -e^{-i \eta _-} \\
\end{array}
\right),
\end{equation}
where $t_{\text{st}}$ is the total duration of a single STIRAP, we took $t_0=0$ without loss of generality, and we used that $U_{\text{det.}}(2t_{\text{st}},t_{\text{st}})=U_{\text{det.}}(t_{\text{st}},0)$ as the two STIRAP processes are assumed the same.
It is evident that probability amplitudes of states $|1\rangle$ and $|3\rangle$ after the interaction are the same as the initial ones, except for a common global phase $(-e^{-i \eta _-})$, thus allowing for robust spin refocusing, e.g., in the presence of amplitude or detuning errors. Embedding STIRAP as a building block for the design of two-photon quantum gates might also be possible, e.g., by combining them with composite pulses, similarly to the design robust adiabatic pulses in two-state systems.

We note that previous work \cite{Yan2018josab} has also discussed exchanging the populations of two ground states in a three-level system by STIRAP but it required very careful adjustment of the single-photon detuning and pulse delay for achieving high fidelity. In contrast, our work (1) demonstrates that swapping is possible for a wide range of single-photon detuning as long as adiabaticity is maintained, (2) it considers swapping the populations of coherent superposition states and statistical mixtures and the differences between these cases, and (3) it shows that while single STIRAP is sensitive to the dynamic phase $\eta_{\pm}$ (see Eq. \eqref{eqPropDetSTIRAP}), it can be applicable for spin refocusing by , e.g., in dynamical docoupling, by using two or an even number of swaps, as demonstrated in Eq. \eqref{eqProp2DetSTIRAP}.

Finally, one can also combine these techniques with other advanced coherent control methods such as optimal control \cite{BoscaiJMP2002,DoriaPRL2011,MuellerRPP2022,Mastroserio2022aqt}, composite pulses \cite{Bruns2018PRA,Levitt1986,Genov2013,Genov2014,Torosov2019pra,Hain2020pra,Torosov2020pra} and shortcuts to adiabaticity \cite{ChenPRL2010,Van-DammePRA2017,OdelinRMP2019} to improve robustness, increase speed and suppress unwanted transitions. Expansion of the methods to multilevel systems is also envisioned.


\section{Conclusion\label{secConclusion}}

We considered the applicability of STIRAP and some of its variants for implementing a robust and efficient population-swapping between two states $|1\rangle$ and $|3\rangle$ in a three-state system. The two states are typically long-lived ground states and it is usually important to minimize the population in the intermediate state $|2\rangle$, which can be fast decaying.
The results show that resonant STIRAP cannot perform swapping successfully but a slight modification, where we introduce a non-zero single-photon detuning, allows for efficient and robust population swapping of the populations of states $|1\rangle$ and $|3\rangle$ for any initial state. When the intermediate state decays fast, it usually preferable to use STIRAP with a large detuning, to avoid populating the lossy state, at the expense of slower implementation. The results of our work could be useful for efficient and robust state preparation, dynamical decoupling and design of quantum gates in ground state qubits via two-photon interactions.


\section*{Acknowledgment\label{secAcknowledgment}}
The authors owe their deep gratitude to B. W. Shore for inspiration and illuminating discussions in many different settings. The authors also acknowledge valuable discussions and useful feedback from Klaas Bergmann, Nikolay V. Vitanov, Junlan Jin, Dominique Sugny, Svetlana Malinovskaya, and Mark Raizen. GG and FJ acknowledge support by DFG, Volkswagenstiftung,  IQST, BMBF Projects 50WM2170, 16KIS1593, 16KIS0832, 13N16215, 13N15440, 03ZU1110JA, DFG Projects  316249678, 445243414, 499424854, and ERC HyperQ Synergy grant 856432. M.A. would like to acknolwlege support from the NATO Science for Peace and Security Programme under grant G5794. The work of DB was supported in part by the Deutsche Forschungsgemeinschaft (DFG, German Research Foundation) Project ID 423116110 and Project ID 390831469:  EXC 2118 (PRISMA+ Cluster of Excellence).

\appendix
\section{Solution for STIRAP with a large detuning by adiabatic elimination\label{secAdiabElimination}}

\begin{figure}[t]
\begin{centering}
	\includegraphics[width=0.8\columnwidth]{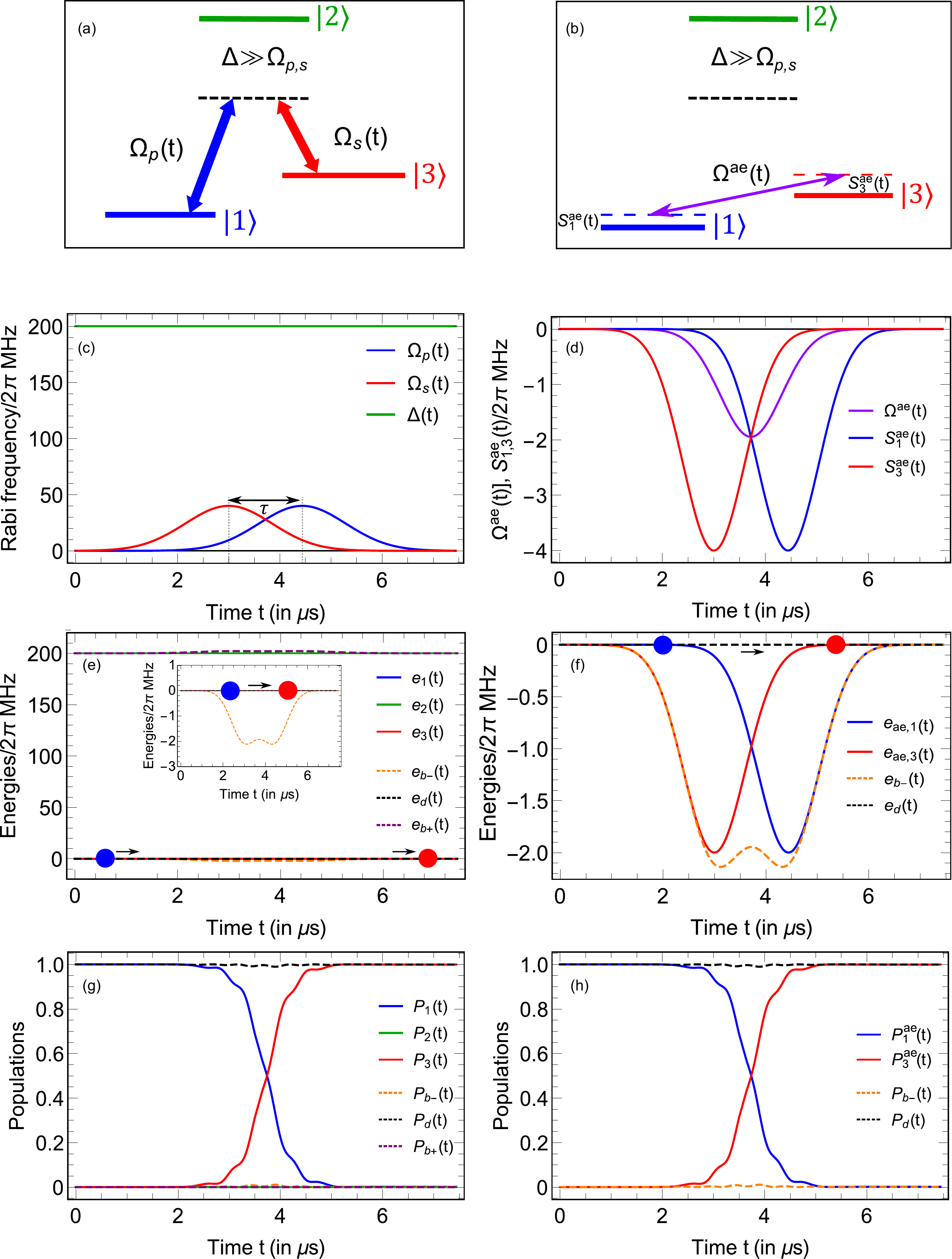}
\caption{(a) Coupling scheme for STIRAP with a large detuning and
(b) the corresponding scheme after adiabatic elimination.
(c) Pulse sequence with the same parameters as in Fig.\,\ref{figdSTIRAP}(b) and
(d) the corresponding two-photon coupling and Stark shifts after adiabatic elimination.
(e) Eigenenergies of the bare and adiabatic states in the full system as in Fig.\,\ref{figdSTIRAP}(c) and
(f) the corresponding eigenenergies in the reduced system, where the adiabatic states $|d\rangle$ and $|b_{-}\rangle$ and are the same for both cases (compare the inset of (c) and (d)). 
The time evolution in the adiabatic basis leads to an avoided crossing, where the bare eigenenergies $\epsilon_{k}^{\text{ae}}(t)$ cross when $\rabi{p}(t)=\rabi{s}(t)$. However, the adiabatic eigenenergies, corresponding to states  $|d\rangle$ and $|b_{-}\rangle$, cannot cross due to the interaction, leading to efficient and robust population transfer.
Numerically simulated time evolution of the populations for (g) the full system as in Fig.\,\ref{figdSTIRAP}(e) and for (h) the reduced system after adiabatic elimination is practically the same. 
	}
	\label{figdSTIRAPae}
\end{centering}
\end{figure}

The case of STIRAP with a large single-photon detuning ${\Delta\gg\rabi{\text{rms}}}(t)$ can also be analyzed by adiabatic eliminatation of the excited state $\ket{2}$ from the system.
The three-state system dynamics are described by the time-dependent Schr\"odinger equation \cite{Shore1990}
 $i\partial_t c(t)=H_\text{RWA}(t)c(t)$, where $c(t)=[c_1(t),c_2(t),c_3(t)]^T$ is a column vector with the probability amplitudes of the three states and the Hamiltonian $H_\text{RWA}(t)$ is given in Eq. \eqref{eqHamiltonianRWA}. We adiabatically eliminate the highly detuned state $|2\rangle$ by taking $\partial_t c_2(t)=0$ and substituting the resulting expression for $c_2(t)$ in the other two equations \cite{Vitanov1997,TorosovJphysB2012,Torosov2013}. Then, the dynamics of the reduced two-state system of states $|1\rangle$ and $|3\rangle$ is described by the Hamiltonian
\begin{align}
\label{eqHamiltonianAE}
H_{\text{ae}}(t)=
\frac{1}{2}
\begin{pmatrix}
S_1^{\text{ae}}(t) & \Omega^{\text{ae}}(t) \\
\Omega^{\text{ae}}(t)^{\ast} & S_3^{\text{ae}}(t)
\end{pmatrix}
=
\frac{\Omega_{\text{rms}}(t)^2}{4\Delta}
\left(
\begin{array}{cc}
 \sin ^2\vartheta (t) & \frac{1}{2}\sin 2\vartheta (t) \\
  \frac{1}{2}\sin 2\vartheta (t) & \cos ^2\vartheta (t) \\
\end{array}
\right),
\end{align}
where $S_1^{\text{ae}}(t)=-|\rabi{p}(t)|^2/(2\Delta)$ and $S_3^{\text{ae}}(t)=-|\rabi{s}(t)|^2/(2\Delta)$ are Stark shifts of states $|1\rangle$ and $|3\rangle$ due to the off-resonant interaction with state $|2\rangle$ and $\Omega^{\text{ae}}(t)=-\rabi{p}(t)\rabi{s}(t)^{\ast}/(2\Delta)$ is the effective coupling between states $|1\rangle$ and $|3\rangle$. In the last equality we assumed that $\Omega^{\text{ae}}(t)$ is real for simplicity of presentation and without loss of generality.
Figures \ref{figdSTIRAPae}(b) and \ref{figdSTIRAPae}(d) show the resulting coupling scheme and temporal behavior of the Rabi frequency and detuning after adiabatic elimination, respectively.
It proves useful to transform our reduced basis by using the rotation matrix
\begin{align}
R^{\text{ae}}(t)&=
\left(
\begin{array}{cc}
 \cos \vartheta (t) & -\sin \vartheta (t) \\
 \sin \vartheta (t) & \cos \vartheta (t) \\
\end{array}
\right),
\end{align}
which allows us to obtain the Hamiltonian in the reduced adiabatic basis after adiabatic elimination
\begin{align}
H^{\text{ae}}_{\text{adiab}}(t)&=R^{\text{ae}}(t) H_\text{ae}(t)R^{\text{ae}~\dagger}(t)-i R^{\text{ae}}(t)\partial_{t}R^{\text{ae}~\dagger}(t)\notag\\
&=\left(
\begin{array}{cc}
 0 & -i \vartheta '(t) \\
 i \vartheta '(t) & -\frac{\Omega _{\text{rms}}(t){}^2}{4 \Delta } \\
\end{array}
\right)
\approx
\left(
\begin{array}{cc}
 0 & 0 \\
 0 & \epsilon_{-} \\
\end{array}
\right),
\end{align}
where we assumed adiabatic evolution in the last equality, i.e., $|\vartheta '(t)|\ll |\epsilon_{-}(t)|$, so we could neglect the effect of the off-diagonal elements of the Hamiltonian, and used that $\epsilon_{-}\approx-\frac{\Omega _{\text{rms}}(t){}^2}{4 \Delta }+O\left(\frac{\Omega _{\text{rms}}(t){}^4}{\Delta^3 }\right)$ in the limit when $\Omega _{\text{rms}}(t)\ll \Delta$ (we assumed $\Delta>0$ without loss of generality).
It is evident that the adiabatic states of the reduced system after adiabatic elimination are $|d\rangle$ and $|b_{-}\rangle$ of the complete system. This is the case as the both the transformation matrix $R^{\text{ae}}(t)$ and the reduced system Hamiltonian $H^{\text{ae}}_{\text{adiab}}(t)$ can be obtained from $R(t)$ in Eq. \eqref{eqRmatrix} and the Hamiltonian $H_{\text{adiab}}(t)$ in Eq. \eqref{eqHamiltonianAdiab}, respectively. We only require the condition $\Omega _{\text{rms}}(t)\ll \Delta$, so the mixing angle $\phi\to 0$, and we consider only the subset of states $|d\rangle$ and $|b_{-}\rangle$ of $H_{\text{adiab}}(t)$ to obtain $H^{\text{ae}}_{\text{adiab}}(t)$. The numerical simulations of the eigenenergies of the adiabatic states of the full and reduced system after adiabatic elimination in the inset of Fig. \ref{figdSTIRAPae}(e) and Fig. \ref{figdSTIRAPae}(f) also confirm that $|d\rangle$ and $|b_{-}\rangle$ are the adiabatic states in both cases.

One can gain additional intuition about the mechanism of adiabatic population swapping with STIRAP with a large detuning by considering the time evolution of the energies of the bare and adiabatic states in the reduced system. The time evolution in the adiabatic basis leads to an avoided crossing, where the bare eigenenergies $\epsilon_{k}^{\text{ae}}(t)=S_{k}^{\text{ae}}(t)/2,~k=1,3$ cross due to the time evolution of their Stark shifts when $\Omega_{p}(t)=\Omega_{s}(t)$, i.e.,  $\vartheta(t)=\pi/4$. However, the adiabatic eigenenergies, corresponding to states  $|d\rangle$ and $|b_{-}\rangle$ approach each other then but cannot cross due to the interaction, see Fig. \ref{figdSTIRAPae}(d). This leads to efficient and robust population transfer. 
Specifically, when the system is initially in $|1\rangle$ it is also in the dark state $|d\rangle$. During the adiabatic evolution it remains in $|d\rangle$ but the latter becomes aligned with $-|3\rangle$ due to the change of the mixing angle $\vartheta(t)$ from $0$ to $\pi/2$. This leads to efficient and robust population transfer (see Fig. \ref{figdSTIRAPae}(e) and Fig. \ref{figdSTIRAPae}(f)). The process is reversed and symmetric when the system is initially in $|3\rangle$ when the population transfer takes place via $|b_{-}\rangle$ and is also efficient and robust. As already shown, successful population transfer from both  $|1\rangle$ and $|3\rangle$ is a sufficient condition for population swapping, which is independent from the initial state. The numerical simulations in Fig. \ref{figdSTIRAPae}(g) and Fig. \ref{figdSTIRAPae}(h) confirm the above analysis and show that the time evolution of populations of the bare and the adiabatic states in both the full and reduced system is the same, demonstrating the validity of the adiabatic elimination approximation.

\section*{References\label{secReferences}}

\bibliographystyle{unsrt}
\bibliography{library}


\end{document}